\documentclass[referee]{raa}            % referee version: for submission

\usepackage{graphicx,times}             %for PS/EPS graphics inclusion, new

\begin{document}

   \title{Long-term photometric behaviour of the eclipsing Z Cam-type dwarf nova AY Psc
%\footnotetext{$*$ Supported by the National Natural Science Foundation of China.}
}
%   \subtitle{I. Place Your Subtitle Here}

   \volnopage{Vol.0 (200x) No.0, 000--000}      %%preserved for Editor. DOn't remove!
   \setcounter{page}{1}          %%starting page, preserved for Editor. DOn't remove!

   \author{Zhong-tao Han
      \inst{1,2,3}
   \and Sheng-bang Qian
      \inst{1,2,3}
   \and Irina Voloshina
      \inst{4}
   \and Li-ying Zhu
      \inst{1,2,3}
   }

   \institute{Yunnan Observatories, Chinese Academy of Sciences (CAS), P. O. Box 110, 650216 Kunming, China; {\it zhongtaohan@ynao.ac.cn}\\
        \and
             Key Laboratory of the Structure and Evolution of Celestial Objects, Chinese Academy of Sciences, P. O. Box 110, 650216 Kunming, China\\
        \and
             University of Chinese Academy of Sciences, Yuquan Road 19\#, Sijingshang Block, 100049 Beijing, China\\
        \and
             Sternberg Astronomical Institute, Moscow State University, Universitetskij prospect 13, Moscow 119992, Russia\\
   }

   \date{Received~~2009 month day; accepted~~2009~~month day}

\abstract{New eclipse timings of the Z Cam-type dwarf nova AY Psc were measured and the orbital ephemeris was revised.
Based on the long-term AAVSO data, moreover, the outburst behaviors were also explored. Our analysis suggests that the normal outbursts are quasi-periodic, with an amplitude of $\sim2.5(\pm0.1)$ mag and a period of $\sim18.3(\pm0.7)$ days. The amplitude vs. recurrence-time relation of AY Psc is discussed, and we concluded that this relation may represents general properties of dwarf nova (DN) outbursts.
The observed standstill ends with an outburst, which is inconsistent with the general picture of Z Cam-type stars. This unusual behavior was considered to be related to the mass-transfer outbursts. Moreover, the average luminosity is brighter during standstills than during outburst cycles. The changes in brightness marks the variations in $\dot{M}_{2}$ due to the disc of AY Psc is nearly steady state. $\dot{M}_{2}$ value was limited to the range from $6.35\times10^{-9}$ to $1.18\times10^{-8}M_{\odot}yr^{-1}$.
More detailed examination shows that there are a few small outbursts presence during standstills. These events with amplitudes of $\sim0.5-0.9$ mag are very similar to the stunted outbursts reported in some NLs. We discussed several possible mechanisms and suggested that the most reasonable mechanism for these stunted outbursts is a changing mass-transfer rate.
}

   \authorrunning{Zhong-tao Han et al. }            %author_head in even pages
   \titlerunning{The Z Cam-type dwarf nova AY Psc}  % title_head in odd pages

   \maketitle
%% The author head (on even pages) and the title head (on odd pages) will be
%% automatically extracted from \author{} and \title{}. Whenever the title is too long,
%% you will be asked to supply a shorter one by inserting either \authorrunning{} or
%% \titlerunning{} before \maketitle. Anyway, you can specify your own heads.
%%
%%
%% Note: In the following text body of your manuscript, please note several differences from
%%       other major journals:
%% (1) \subsection{Please Capitalize the First Letter of Each Notional Word in Subsection Title}
%% (2) Please Capitalize the First Letter of Each Notional Word in all tables' captions

%
%________________________________________________ sections below
%
\section{Introduction}           %% first-level sections will be auto-capitalized
\label{sect:intro}

Z Cam stars are a subtype of dwarf nova-type (DN) cataclysmic variables (CVs) exhibiting protracted standstills about 0.7 mag below maximum brightness, during which the brightness stays constant (Warner 1995). The duration of standstills has a large range from tens of days to years. These stars exhibit U Gem-type outbursts but the standstills are similar to the behaviour of nova-like CVs (NLs). Osaki (1974) proposed an interpretation of standstills for the first time in which the accretion disc was in the steady stage. After that, these systems are defined as an intermediate between stable NLs and unstable DNe (Smak 1983). Meanwhile, a model for the standstills was presented by Meyer \& Meyer-Hofmeister (1983). The model suggested that the mass transfer rate in Z Cam stars is close to a critical rate (i.e. $\dot{M}_{2}\approx\dot{M}_{crit}$). The system with $\dot{M}_{2}<\dot{M}_{crit}$ will generate U Gem-type normal outbursts. While when $\dot{M}_{2}>\dot{M}_{crit}$, the system enters standstill behaving like NLs. Moderate fluctuation in $\dot{M}_{2}$ could produce Z Cam-type light curves (Lin et al. 1985; Buat-M\'{e}nard et al. 2001). Changes in $\dot{M}_{2}$ was generally explained as the irradiation effects of the secondary star's surface by accretion disk (Meyer \& Meyer-Hofmeister 1983). As expected, the average brightness in NLs was higher (by $\sim3$ mag) than in DNe at the same orbital period (Warner 1995). On long time scales, the average brightness in Z Cam stars are expected to follow changes in $\dot{M}_{2}$ because their disks are nearly steady state.

AY Psc was classified as a Z Cam star due to its occasional standstills and normal DN outbursts (Mercado \& Honeycutt 2002). It was identified as a CV candidate by Green et al. (1982) and first studied in detail by Szkody et al. (1989) who presented $B-$ and $V-$band light curves showing deep eclipses with an orbital period of 5.13 h. Later, an orbital ephemeris was provided by Diaz \& Steiner (1990) who also revised the orbital period as 5.2 h. The system parameters were derived by Howell \& Blanton (1993) using photometric analysis and by Szkody \& Howell (1993) using the time-resolved spectroscopy. According to Szkody \& Howell (1993), AY Psc contains a massive $\sim1.31M_{\odot}$ white dwarf and a $\sim0.59M_{\odot}$ companion,  for
a total system mass of $\sim1.9M_{\odot}$, which is well in excess of the
Chandrasekhar limit ($\sim1.4M_{\odot}$) and therefore in the parking lot for potential SN Ia
progenitors.
However, we still know little about its outbursts because of the paucity of the long-term photometric data. Fortunately, many V-band data from American Association of Variable Star Observers (AAVSO) covering a timescale of $\sim6$ yr provide a good opportunity to ascertain the outburst properties. Moreover, the deeply eclipsing nature of AY Psc can be used to probe the orbital period changes. From an evolutionary perspective, the period analysis can offer some clues concerning the orbital evolution and circumbinary companions.
The most common method for determining the period variations is to analyze the observed-calculated ($O-C$) diagram. We have used this method several times in the past few years to study a few eclipsing CVs such as Z Cha (Dai et al. 2009), OY Car (Han et al. 2015), V2051 Oph (Qian et al. 2015) and GSC 4560-02157 (Han et al. 2016).
In this paper, we present new CCD photometric observations of AY Psc and update the orbital ephemeris. Then its outburst properties were analysed by using the AAVSO data.

%% Authors can give a citation as 'Michel et al. 1992'.
%% You may also use \cite, \citep and \citet for citation, and use Table~1 or Figure~1
%% and so forth. Using \ref and \label for cross-references of Tables/Figures
%% is a good way in adjusting/adding/removing text, tables or figures.

\section{Observations and Data preparation}
\label{sect:Obs}
New CCD photometric observations of AY Psc were carried out by using several different telescopes. They were: the 60 cm and the 1.0 m reflecting telescopes mounted an Andor DW436 2K CCD camera at Yunnan Observatories (YNO); the 85 cm and the 2.16 m telescopes at the XingLong station of the National Astronomical Observatory (NAO); and the 2.4 m Thai National Telescope (TNT) of National Astronomical Research Institute of Thailand (NARIT). The 85-cm telescope was mounted an Andor DW436 1K CCD camera and the 2.16 m telescope were equipped with a PI $1274\times1152$ TE CCD. The TNT is a Ritchey-Chr\'{e}tien with two focuses and an ULTRASPEC fast camera was attached on it. During the observations, no filters were used in order to improve the time resolution. The aperture photometry package of IRAF was used to reduce the observed CCD images. Differential photometry was performed, with a nearby non-variable comparison star. Fig. 1 displays four eclipsing profiles observed with different telescopes. As done by Diaz \& Steiner (1990), new mid-eclipse times are determined by using a cubic polynomial fitting. The errors are the standard deviation values. All mid-eclipse times and the errors are listed in Table 1.

To investigate the outburst properties, the long-term light curves were required. Fig. 2 shows the entire AAVSO light curve of AY Psc in $V-$ and $R-$ band from 2010 August 19$-$2016 November 29. Here we only used the digital photometric data because the errors in the visual data are too large ($\sim0.3-0.5$ mag). The digital photometric observations began in August, 2010 and over $1400$ observations from 26 observers were covered in Fig. 2.
The errors were estimated to within $0.1$ mag for the bright state (during outburst and standstill) and within $0.2$ mag for the quiescence using the comparison and check stars of known magnitude in this field by these observers. Due to AY Psc is an eclipsing system, such data have been prepared by omitting the data during eclipse before the analysis. Although the fragmentary data and the gaps present, the standstill is clearly visible in the light curve. The long-term light curve was partitioned into seven data sets by the gaps. The best data occur after JD $2457553$, where the density of observations is $1.7$ day$^{-1}$ and every day contains at least one data point. Continuous outbursts in this data set were plotted in Fig. 4 and reveal the feature similar to sine wave. More detailed analysis was given in Section 3.2.

\begin{figure}[!h]
\begin{center}
\includegraphics[width=1.0\columnwidth]{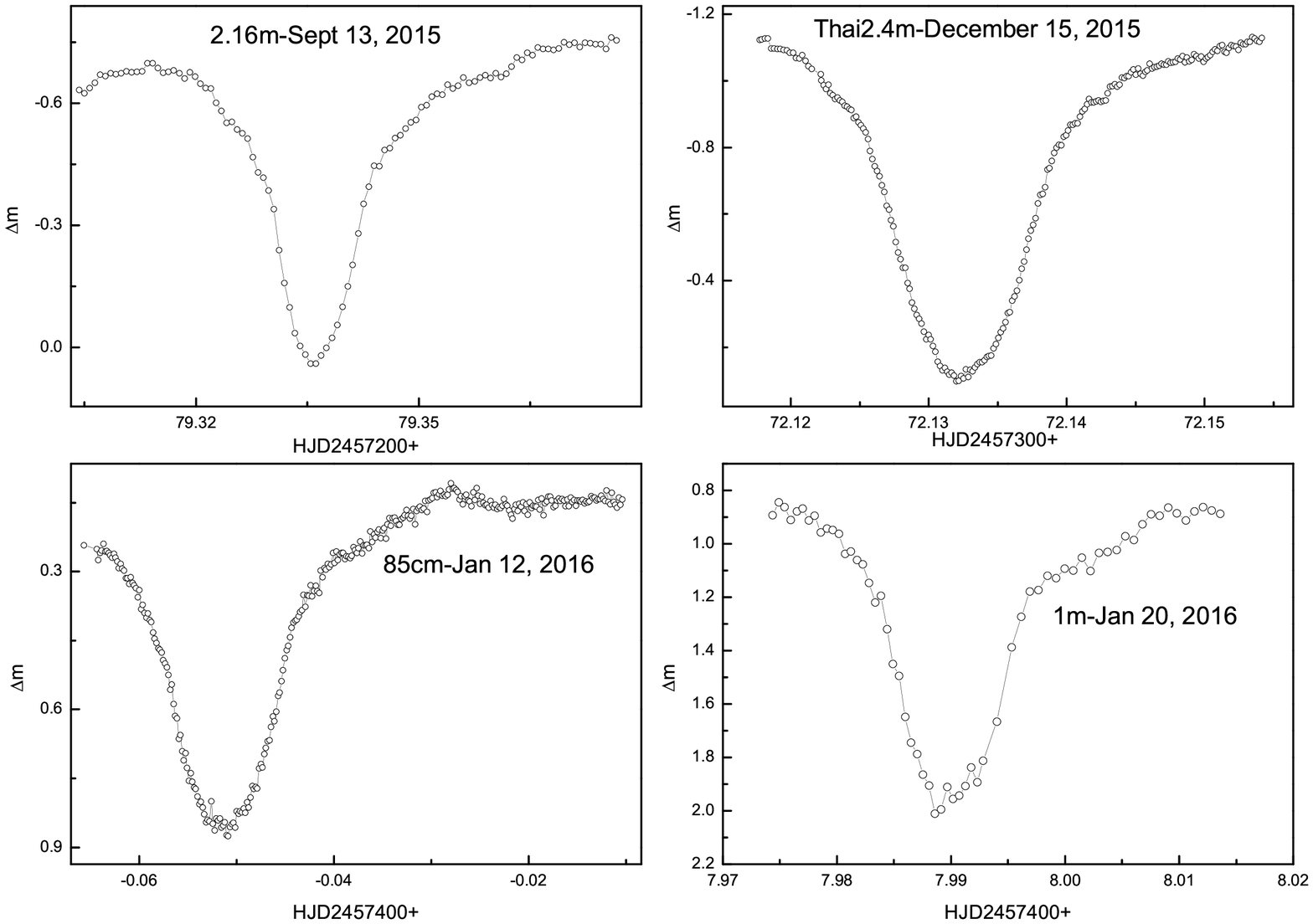}
\caption{Four eclipsing profiles of AY Psc obtained with four different telescopes in China and in Thailand.}
\end{center}
\end{figure}

\begin{figure}[!h]
\begin{center}
\includegraphics[width=0.9\columnwidth]{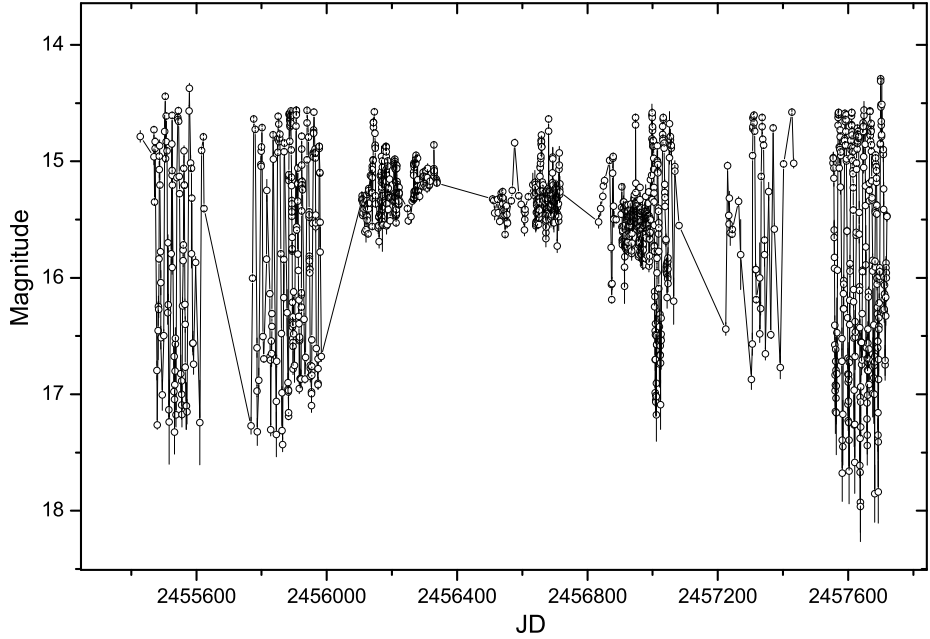}
\caption{The long-term light curve of AY Psc from AAVSO observations during 2010 August$-$2016 November. Note the presence of obvious standstills at the $15.3(\pm0.2)$ mag.}
\end{center}
\end{figure}

\begin{table*}[!h]
\caption{New CCD mid-eclipse times of AY Psc.}
 \begin{center}
 \small
   \begin{tabular}{llcllll}\hline\hline
Date                    & Min.(HJD)        &  E          & $O-C$      & Errors   & Filters  &Telescopes    \\\hline
2012 Dec 13             &2456275.09576     &39811        &-0.01289    &0.00011   &  N  &60cm       \\
2015 Sep 13             &2457279.33629     &44432        &-0.01224    &0.00005   &  N  &2.16m      \\
2015 Sep 24             &2457290.20122     &44482        &-0.01335    &0.00019   &  N  &1m         \\
2015 Dec 15             &2457372.13246     &44859        &-0.01209    &0.00003   &  N  &Thai-2.4m  \\
2016 Jan 12             &2457399.94876     &44987        &-0.01287    &0.00003   &  N  &85cm       \\
2016 Jan 20             &2457407.98993     &45024        &-0.01257    &0.00008   &  N  &1m         \\
\hline
\end{tabular}
\end{center}
\end{table*}

\begin{figure}
\begin{center}
\includegraphics[width=0.9\columnwidth]{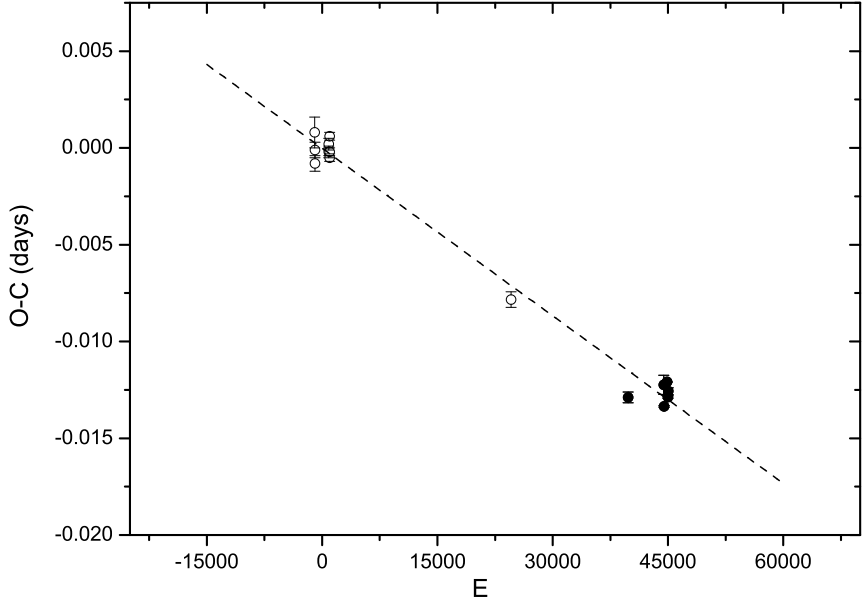}
\caption{$O-C$ diagram of AY Psc. The open circles and solid circles denote the data in literature and our observations, respectively. The dashed line represents the best-fitting linear ephemeris.}
\end{center}
\end{figure}

\section{Results and Discussion}
\subsection{Revised ephemeris}
Mid-eclipse times of AY Psc have been published in the literatures by two authors (e.g. Diaz \& Steiner 1990 and G\"{u}lsecen et al. 2009). Diaz \& Steiner (1990) reported 9 mid-eclipse times and  showed no sign of any orbital period change. Combining the previous data with our observations, the latest version of $O-C$ diagram was displayed in Fig. 3. The $O-C$ values of all available data were computed with the linear ephemeris given by Diaz \& Steiner (1990):
\begin{equation}
Min.I = HJD\,2447623.3463+0.2173209\times{E},
\end{equation}
where HJD\,2447623.3463 is the initial epoch and 0.2173209 d is orbital period. The best-fitting linear ephemeris to all the eclipse times of AY Psc is
\begin{equation}
Min.I = HJD\,2447623.34628(2)+0.21732061(1)\times{E}.
\end{equation}
The dashed line in Fig. 3 represents the linear revised ephemeris. However, the residuals from this ephemeris show some deviations and the orbital period does not seem to be constant.
This may be caused by some of the unknowns or a true period change. But there is still no more evidence to support this change.
Therefore, further observations are critically required to ascertain the changes in the orbital period of this system.

\subsection{Outburst properties}

\subsubsection{Normal Outbursts and K$-$P relation}

In Fig. 2 we show the long-term AAVSO light curve of AY Psc during 2010 August$-$2016 November. The data were divided into seven segments because a lot of data are missing.
A detailed view of the light curves reveals three main features: normal outburst, standstill and stunted outburst. During normal outburst, the duty cycle of AY Psc approaches $100\%$. This behavior is thought to be due to $\dot{M}_{2}\approx\dot{M}_{crit}$ (Lin, Papaloizou \& Faulkner 1985; Warner 1995). The system rises from $V\approx17.1$ at minimum to $V\approx14.6$ at maximum in $6-8$ day, exhibiting a mean outburst amplitude of $\sim2.5(\pm0.1)$ mag.
Fig. 4 displays a best outburst data set for AY Psc covering 169 days (2016 Jun 14$-$2016 Nov 29). A sine curve fit was made to the eight outburst cycles in Fig. 4, giving a recurrence time of $\sim19.03(\pm0.04)$ days. However, the full light curve contains at least four outburst segments. To verify this result, the analysis of all outburst segments is required. Fig. 5 shows all outburst data sets and corresponding power spectrums. The power spectrums correspond to the periods between 17.7 and 18.5 days, which is smaller than the sine-fitting period ($\sim19.03$ days). This implies that the outburst in AY Psc is not strictly periodic but quasi-periodic.

\begin{figure}[!h]
\begin{center}
\includegraphics[width=0.9\columnwidth]{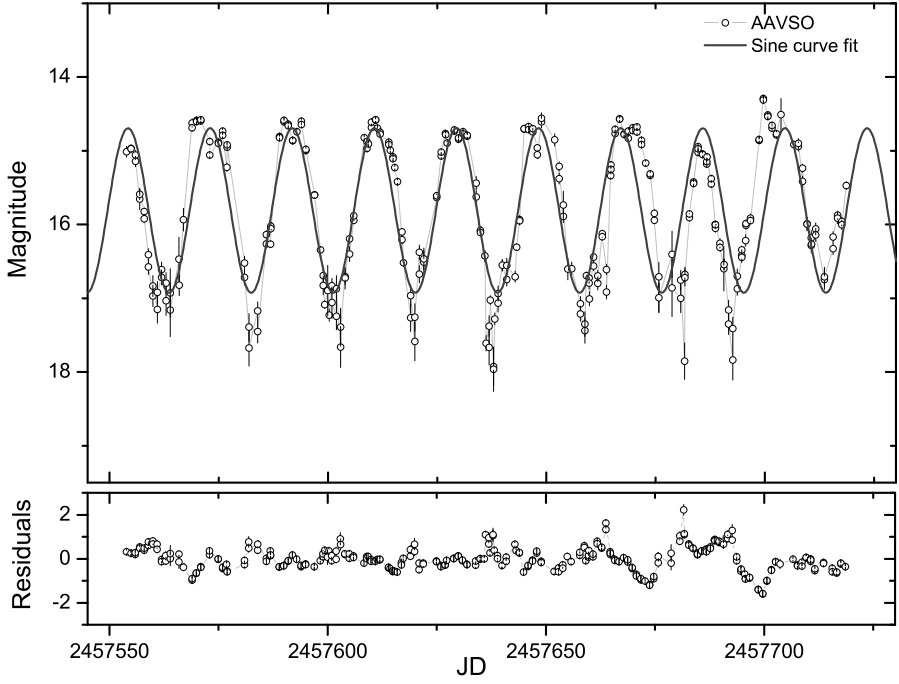}
\caption{Continuous outbursts in AY Psc spanning $\sim170$ days (2016 Jun 14$-$2016 Sept 27). The sine curve fit was superposed on this outburst segment, which gives a period of $\sim19.03(\pm0.04)$ days. The lower panel displays the fitting residuals from
the sine wave.}
\end{center}
\end{figure}

\begin{figure}
\begin{center}
\includegraphics[width=1.0\columnwidth]{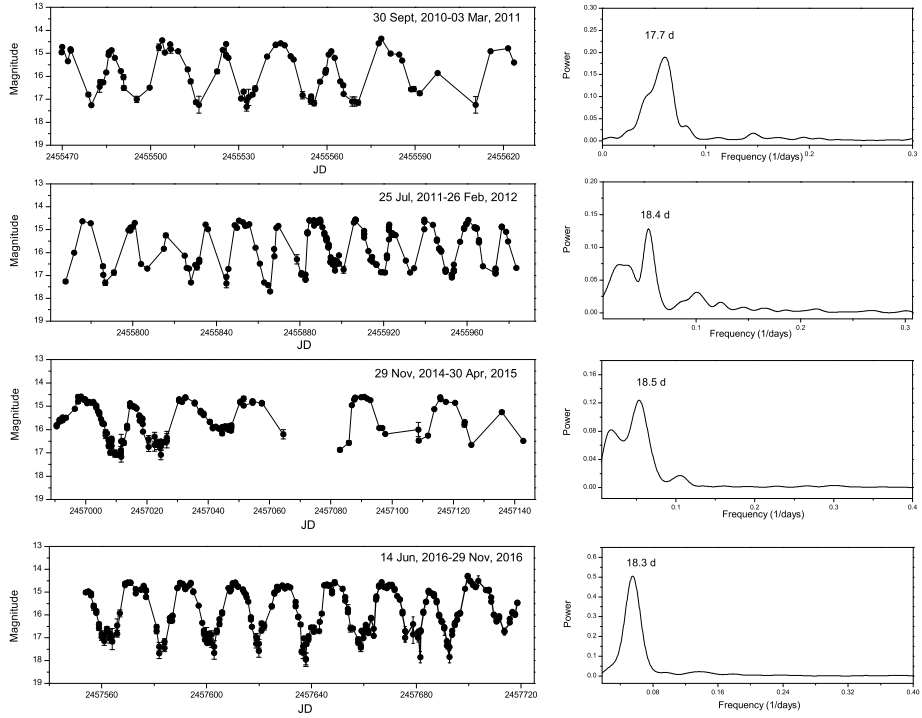}
\caption{Four outburst segments and corresponding power spectrums. The power spectrums indicate that the outbursts are quasi-periodic, with  the periods between 17.7 and 18.5 days.}
\end{center}
\end{figure}

\begin{figure}[!h]
\begin{center}
\includegraphics[width=1.0\columnwidth]{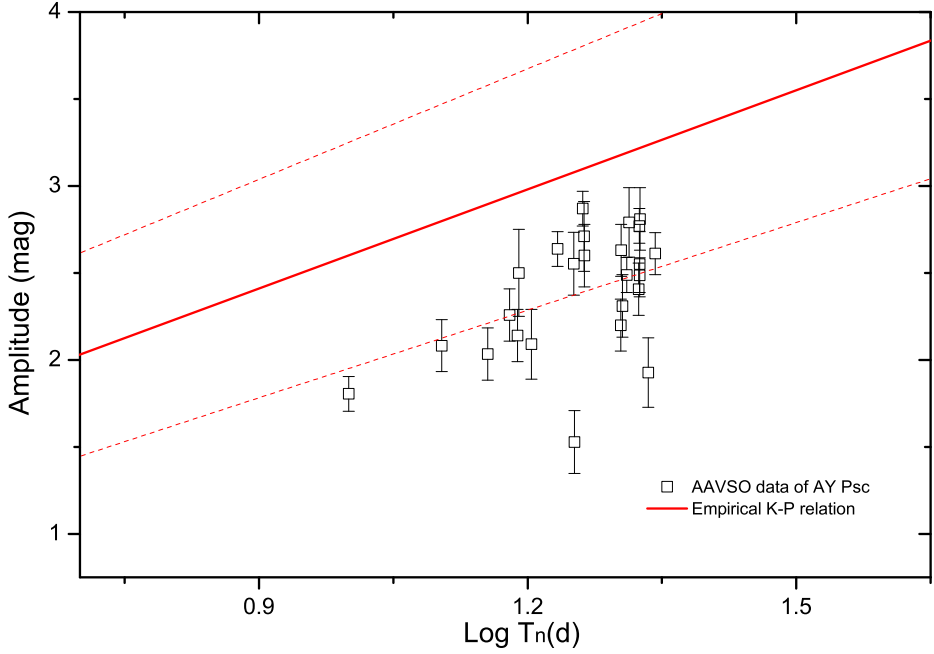}
\caption{Comparison of the K-P relation and AAVSO data for AY Psc. The open squares represent the statistical data, the red solid line refer to empirical K-P relation (Warner 1995) and red dashed lines denote its upper and lower uncertainties.}
\end{center}
\end{figure}

\begin{figure}
\begin{center}
\includegraphics[width=0.9\columnwidth]{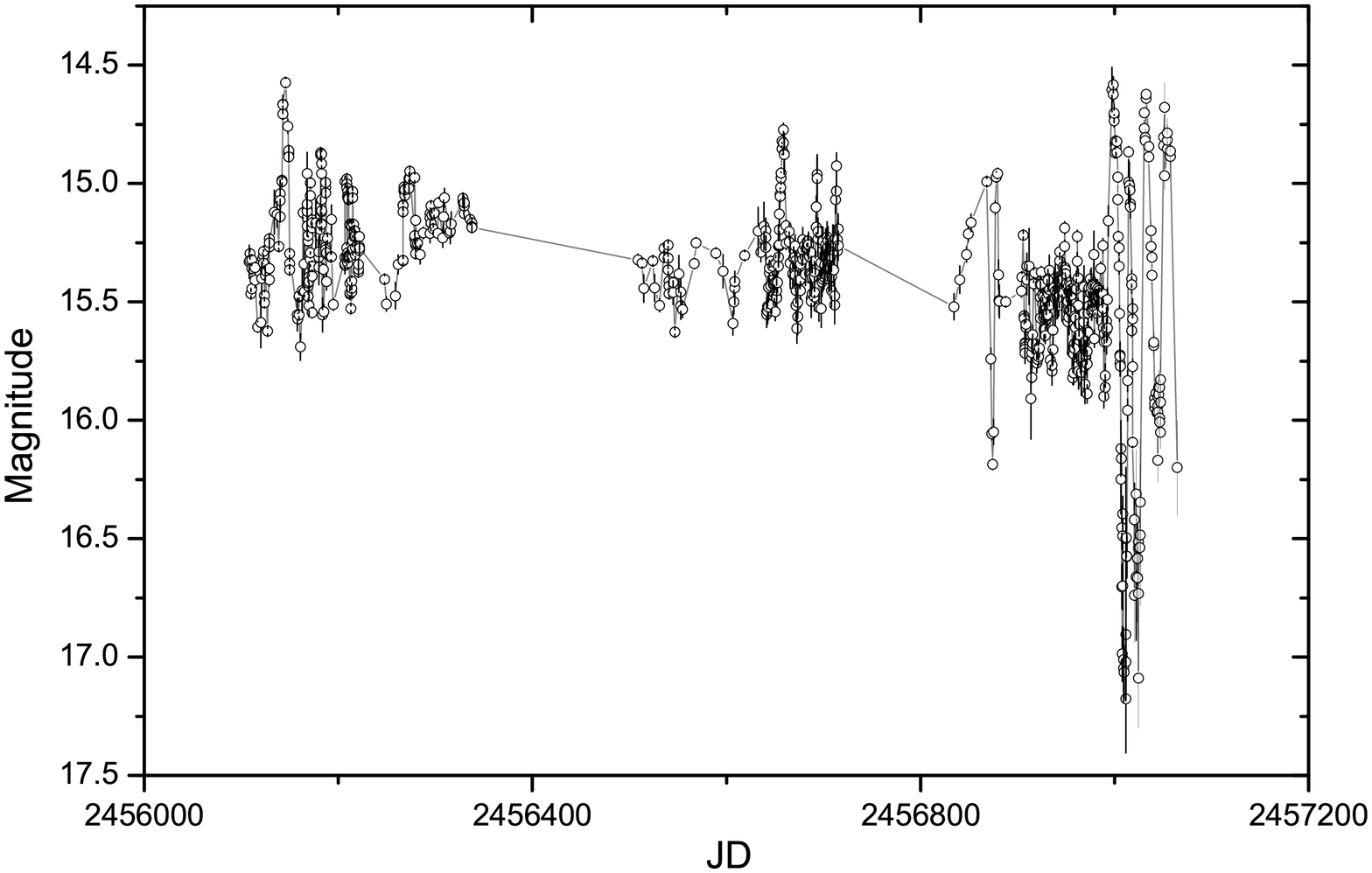}
\caption{The standstill data at the $15.3(\pm0.2)$ mag extracted from Fig. 2 during the period from 2012 June 29 to 2015 January 1.
It is clear that there are small amplitude oscillations present during standstills.}
\end{center}
\end{figure}

Kukarkin \& Parenago (1934) first noted that there was a relation between the outburst amplitude ($A_n$) and the outburst recurrence time ($T_n$) for DN and recurrence nova (RN). After that, this relation has done many times revised and improved to constrain its application range. Finally, a general correlation was found by analysing DN normal outbursts (van Paradijs 1985). The most recent version of K-P relation from Warner (1995) is as follows:
\begin{equation}
A_n=0.7(\pm0.43)+1.9(\pm0.22)\log T_n.
\end{equation}
This is an empirical relation. To explore if this relation is model-dependent, moreover,  a theoretical K-P relation was derived by Kotko \& Lasota (2012) using the disc instability model (DIM):
\begin{equation}
A_n=C_1+2.5\log T_n,
\end{equation}
where the constant term $C_1=2.5\log 2\tilde{g}-2.5\log t_{dec}+BC_{max}-BC_{min}$, which depends on the properties of white dwarf and the viscosity parameter $\alpha$.
The outburst parameters of AY Psc and the K-P relation are plotted in Fig. 6. The open squares represent the statistical data from AAVSO database, the red solid line refer to empirical K-P relation and red dashed lines denote its upper and lower uncertainties. However, the observational data were not covered in this relation. Note that this relation for only those systems with $A_{n}>2.5$ mag is significant (Warner 1995). The lower uncertainty of the empirical relation follows the overall trend of observational data reasonably well because the amplitude of AY Psc falls on the boundary of $2.5$ mag. Therefore, the K-P relation for AY Psc will be roughly replaced by the lower limit of the empirical relation
\begin{equation}
A_n=0.27+1.68\log T_n.
\end{equation}
Based on those discussions, we suggest that the K-P relation may represent the common nature of DN outbursts.

\subsubsection{Standstills and brightness modulation}

Fig. 7 shows the standstill signal at the $15.3(\pm0.2)$ mag extracted from Fig. 2 during the period from 2012 June 29 to 2015 January 1. The upper panel of Fig. 8 displays a well-observed transition from standstill to outburst, where we see an unusual characteristic behavior being differ from most manner of Z Cam stars. This standstill ends with an outburst (see Fig. 8). However, the model of Z Cam stars predicts that a standstill can only be terminated by a decline to quiescence rather than an outburst (Buat-M\'{e}nard et al. 2001).
Before several the unusual Z Cam systems were discovered, only one recorded standstill in the prototype star Z Cam terminates in a rise to maximum (Warner 1995; Oppenheimer et al. 1998). Recently,  this rare behavior is observed in several anomalous Z Cam-type stars such as AH Her, IW And, V513 Cas and ST Cha (Simonsen 2011; Simonsen et al. 2014). A plausible explanation for this behavior was proposed by Hameury \& Lasota (2014), who indicated that the mass-transfer outbursts can reproduce the observed properties of these unusual systems. This implies that the changes in mass transfer from the secondary star are responsible for the changes in brightness on longer timescales. For Z Cam-type systems, a direct test is to compare the standstill and outburst brightness.
In effect, there have been several studies to do this work (Lortet 1968; Oppenheimer et al. 1998; Honeycutt et al. 1998). To clearly reveal the mass-transfer variations, the mean brightness during standstills and the outbursts were calculated by averaging long-term AAVSO data. In computing the mean magnitude values, only full cycles were averaged. For outbursts, two mean values were measured per outburst cycle, overlapping by one-half cycle. For standstills, means were computed for successive intervals of several ten days. The mean brightness values of all data are plotted in Fig. 9.
The red dots in Fig. 9 represent the means during standstills and the black dots denote the means during outbursts.
Clearly, the average magnitudes during standstills is brighter than during outburst intervals. This result is consistent with the previous authors (e.g. Oppenheimer et al. 1998; Honeycutt et al. 1998). Moreover, the trend in Fig. 9 also means the changes in $\dot{M}_{2}$, in agreement with the general picture of the standstills. The green solid line marks the transition between outburst cycles and standstills, corresponding to the $\dot{M}_{crit}$. Frank et al. (2002) give an expression for $\dot{M}_{crit}$:
\begin{equation}
\dot{M}_{crit}\simeq3\times10^{-9}(P_{orb}/{3})^2 M_{\odot}yr^{-1}
\end{equation}
where $P_{orb}$ is in hours. For AY Psc, $P_{orb}=5.22$ h, we find $\dot{M}_{crit}\simeq9.07\times10^{-9}M_{\odot}yr^{-1}$. Buat-M\'{e}nard et al. (2001) pointed out that $\dot{M}_{2}$ will vary by about $30\%$ around $\dot{M}_{crit}$. More specifically, $\dot{M}_{2}$ in AY Psc should be restricted to the range from $6.35\times10^{-9}$ to $1.18\times10^{-8}M_{\odot}yr^{-1}$.

\begin{figure}
\begin{center}
\includegraphics[width=0.9\columnwidth]{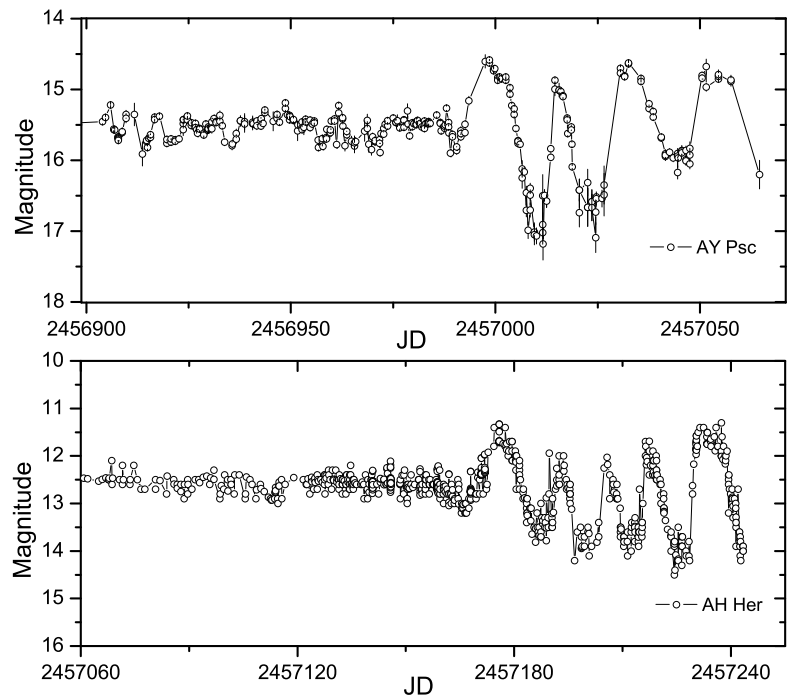}
\caption{Top panel: a transition from standstill to outburst interval in AY Psc. Lower panel: a transition from standstill to outburst interval in AH Her. The two transitions terminate in an outburst, in disagreement with the current theoretical prediction and observations.}
\end{center}
\end{figure}

\begin{figure}
\begin{center}
\includegraphics[width=0.9\columnwidth]{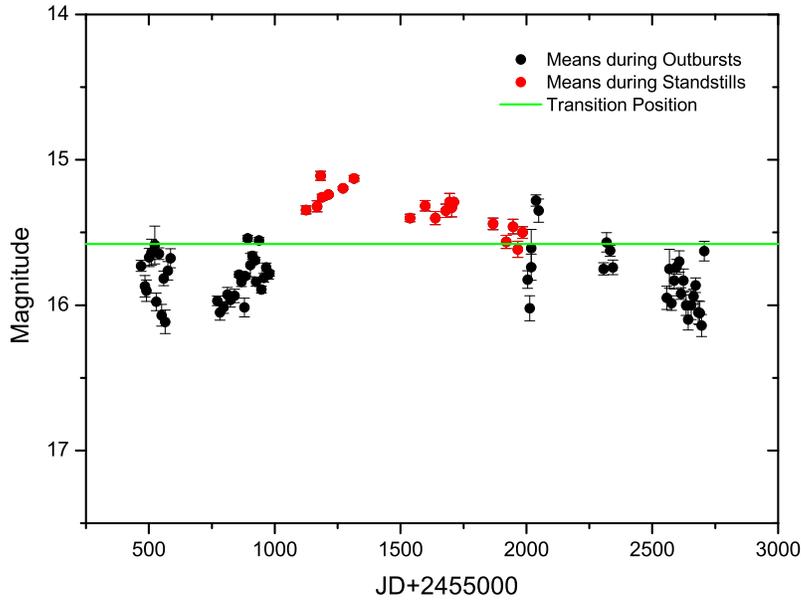}
\caption{Comparison of the outburst and standstill mean brightness in AY Psc using all AAVSO data. The red dots are the means during standstills and the black dots denote the means during outburst cycles. The green line represents the switched position.}
\end{center}
\end{figure}

\subsubsection{Stunted outbursts in AY Psc}

More detailed inspection for the AAVSO data reveals that there are diverse behaviors during standstill. Note that the brightness of the standstill is not constant, but keep fluctuating.
The oscillations with an amplitude of $\sim0.2$ mag can be seen by eye in Fig. 7 and 8. Moreover, we find occasional outburst-like events during standstill, shown in Fig. 10. These events are best characterized as the small amplitude outbursts, which have properties very similar to the stunted outbursts seen in some NLs (e.g. Honeycutt et al. 1995, 1998, 2001, 2014; Warner 1995; Hoard et al. 2000;  Ramsay et al. 2016). Table 2 summarizes the properties of stunted outbursts in AY Psc. The stunted outbursts in Fig. 10 are visible with amplitudes of $\sim0.5-0.9$ mag and FWHM of $\sim2-14$ days. The rises are slower than the declines, implying that the outbursts are Type B (inside-out).
However, to date, the origin of stunted outbursts is still uncertain. Some possible mechanisms are proposed to account for the stunted outbursts, including a disk truncated by the magnetic field of the white dwarf or a very hot white dwarf, dwarf nova-type outbursts being related to the standard disc instability and the mass transfer modulations (Honeycutt et al. 1998, 2001, 2014; Ramsay et al. 2016). For AY Psc, a truncated disk mechanism can rule out because the inside-out outbursts (Type B) in a truncated disk are infrequent (Honeycutt et al. 1998). During standstill, $\dot{M}_{2}$ from the secondary star exceeds $\dot{M}_{crit}$, the disk stays in the steady state, without DN outbursts. In addition, Warner (1995) indicated that dwarf nova-type outbursts have faster rise times than decline times by a factor of $\sim2$ (i.e. Type A outbursts), which is not compatible with the stunted behaviors of AY Psc. Therefore, the disc instability may not be cause of the stunted outbursts. It seems that a changing mass transfer is a reasonable candidate mechanism for these stunted behaviors. First, the steady-state disc allows the brightness to follow changes in mass transfer. Second, Hameury \& Lasota (2014) have studied the disc response to a mass transfer outburst during standstills and reproduced an outburst with an amplitude of $\sim0.8$ mag started from a steady state disc. Meanwhile, the possible origins of the mass transfer outbursts were also discussed by these authors. Nevertheless, these stunted outbursts are not well understood so far.

\begin{figure}
\begin{center}
\includegraphics[width=1.0\columnwidth]{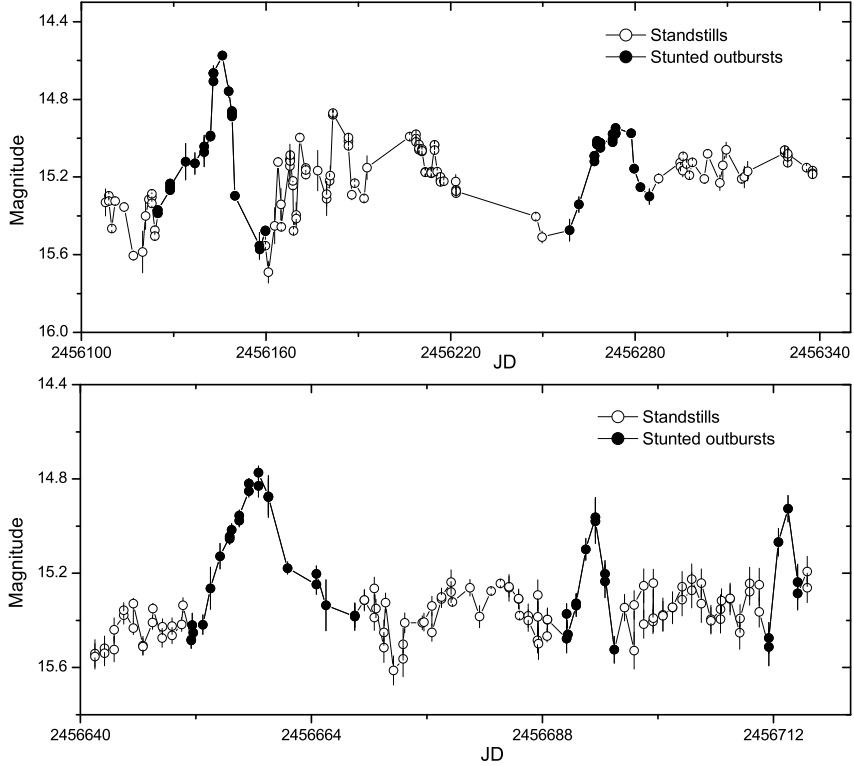}
\caption{Five stunted outbursts during standstills showing the amplitudes of $\sim0.5-0.9$ mag. The characteristics of the stunted outbursts were summarized in Table 2.}
\end{center}
\end{figure}

\begin{table*}[!]
\caption{Basic parameters of the stunted outbursts in AY Psc. }
\begin{center}
 \small
   \begin{tabular}{cccccc}\hline\hline
Outburst  \qquad   &JD   \qquad  &Ampl./mag \qquad  &FWHM/days \qquad  &$\tau_{rise}$/days \qquad  &$\tau_{fall}$/days \\\hline
1  \qquad   &2456145 \qquad  &0.89  \qquad  &7.3 \qquad  &21 \qquad   &10\\
2  \qquad   &2456273 \qquad  &0.53  \qquad  &14.1 \qquad  &15 \qquad   &11\\
3  \qquad   &2456658 \qquad  &0.68  \qquad  &6.5 \qquad  &8 \qquad   &10\\
4  \qquad   &2456693 \qquad  &0.56  \qquad  &2.8 \qquad  &3 \qquad   &2\\
5  \qquad   &2456713 \qquad  &0.59  \qquad  &2.1 \qquad  &2 \qquad   &1\\\hline
\end{tabular}
\end{center}
\end{table*}

\section{Summary}

We have presented the photometric results of the eclipsing Z Cam-type star AY Psc using our observations together with AAVSO data.
Our analysis is focus on the orbital period and outburst properties.
The main conclusions of this paper can be summarized as follows.

(i) The orbital period has been revised from 0.2173209 d to 0.21732061 d.

(ii) The duty cycle of AY Psc is close to $100\%$.

(iii) The observed standstill, extending from June 29, 2012 to December 1, 2014, ended with
an outburst rather than a decline to quiescence, as rarely seen in similar systems.

(iv) During the standstill, several stunted outbursts were present.

\begin{acknowledgements}
This work is supported by the Chinese Natural Science Foundation (Grant No. 11133007 and 11325315), the Strategic Priority Research Program ''The Emergence of Cosmological Structure'' of the Chinese Academy of Sciences (Grant No. XDB09010202) and the Science Foundation of Yunnan Province (Grant No. 2012HC011). We thank the numerous observers worldwide who contributed the over 1400 observations of AY Psc to the AAVSO database that made this work possible. Finally, we thank the anonymous referee for those helpful comments and suggestions.
\end{acknowledgements}

\label{lastpage}


\begin{thebibliography}{99}
%% you can type \apj for ApJ, \aap for A&A, \apss for Ap&SS, etc. Please consult
%% the macro chjaa.cls. You can also find them in aasguide.tex (AASTeX for ApJ, AJ, PASP)
%% Please follow the format of ChJAA's reference list

\bibitem[Buat (2001)]{2001....366..612B} Buat-M{\'e}nard, V., Hameury, J.-M., \& Lasota, J.-P. \ 2001, \aap, 366, 612
\bibitem[Dai (2009)]{2009AAS....131..119C} Dai, Z. B., Qian, S. B., Fern{\'a}ndes Laj{\'u}s, E., \ 2009, \apj, 703, 109
\bibitem[Diaz \& Steiner (1990)]{1990....238..170D} Diaz, M.P., \& Steiner, J.E., \ 1990. \aap, 238, 170
\bibitem[Frank (2002)]{2002....128F} Frank J., King A., Raine D., \ 2002, Accretion Power In Astrophysics, Cambridge University Press, 128
\bibitem[Green et al. (1982)]{1982....94..560G} Green, R.F., Ferguson, D.H., et al. \ 1982, PASP, 94, 560
\bibitem[G\"{u}lsecen et al. (2009)]{2009....14..330G} G\"{u}lsecen, H., Retter, A., Liu, A., \& Eseno\v{g}lu, H., \ 2009, New A, 14, 330
\bibitem[Hameury \& Lasota (2014)]{2014....569..48H} Hameury, J. M., \& Lasota, J. P., \ 2014, \aap, 569, 48
\bibitem[Han (2015)]{2015....128H} Han, Z.-T, Qian, S. B.,Fern{\'a}ndes Laj{\'u}s, E. et al. 2015, New A, 34, 1
\bibitem[Han (2016)]{2016....128H} Han, Z.-T, Qian, S. B., Irina Voloshina. et al. 2016, RAA, 16, 156
\bibitem[Hoard et al. (2000)]{2000....112..1595H} Hoard, D. W., Szkody, P., Honeycutt R. K., et al. \ 2000, PASP, 112, 1595
\bibitem[Honeycutt (1995)]{1995....28P} Honeycutt R. K., Robertson J. W., Turner G. W., \ 1995, \apj, 446, 838
\bibitem[Honeycutt (1998)]{1998....28P} Honeycutt R. K., Robertson J. W., Turner G. W., \ 1998, \aj, 115, 2527
\bibitem[Honeycutt (2001)]{2001....473P} Honeycutt R. K., \ 2001, PASP, 113, 473
\bibitem[Honeycutt (2014)]{2014....10P} Honeycutt R. K., Kafka S., Robertson J. W., \ 2014, \apj, 147, 10
\bibitem[Howell \& Blanton (1993)]{1993....106..311H} Howell S. B., \& Blanton S. A., \ 1993, \aj, 106, 311
\bibitem[Kotko \& Lasota (2012)]{2012....28K} Kotko, I., \& Lasota, J.P., \ 2012, \aap, 545, 115
\bibitem[Kukarkin \& Parenago (1934)]{1934....28P} Kukarkin, B. W., \& Parenago, P. P., \ 1934, Var. Star. Bull., 4, 44
\bibitem[Lin et al. (1985)]{1985....212..105L} Lin, D. N. C., Papaloizou, J., \& Faulkner, J. \ 1985, \mnras, 212, 105
\bibitem[Lortet (1968)]{1968....4..381L} Lortet, M.-C. \ 1968, in IAU Colloq. 4, Variable Stars, Nonperiodic Phenomena in Variable Stars, ed. L. Detre (Budapest: Academic Press), 381
\bibitem[Mercado et al. (2002)]{2002....201..40M} Mercado, L., \& Honeycutt, R.K., \ 2002, AAS 201, 40
\bibitem[Meyer et al. (1983)]{1983....121..29M} Meyer, F., \& Meyer-Hofmeister, E. \ 1983, \aap, 121, 29
\bibitem[Oppenheimer et al. (1998)]{1998....115..1175O} Oppenheimer, B. D., Kenyon, S. J., \& Mattei, J. A. \ 1998, \aj, 115, 1175
\bibitem[Osaki (1974)]{1974PASJ....26..429O} Osaki, Y. \ 1974, PASJ, 26, 429
\bibitem[Qian et al. (2015)]{2015APJS....221..17P} Qian, S.-B., Han, Z.-T., et al. \ 2015, \apjs, 221, 17
\bibitem[Ramsay (2016)]{2016....28P} Ramsay G., Hakala P., Wood M. A., Howell s. b. et al., \ 2016, \mnras, 455, 2772
\bibitem[Simonsen (2011)]{2011....39..66S} Simonsen, M. \ 2011, J. Am. Assoc. Var. Star Obs. (JAAVSO), 39, 66
\bibitem[Simonsen et al. (2014)]{2014....42..199S} Simonsen, M., Bohlsen, T., Hambsch, F.-J., \& Stubbings, R. \ 2014, JAAVSO, 42, 199
\bibitem[Smak (1983)]{1983....272..234S} Smak, J. \ 1983, \apj, 272, 234
\bibitem[Szkody et al. (1989)]{1989....101..899S} Szkody, P., Howell, S. B., et al. \ 1989, PASP, 101, 899
\bibitem[Szkody et al. (1993)]{1993....403..743S} Szkody P., \& Howell S. B., \ 1993, \apj, 403, 743
\bibitem[van Paradijs (1985)]{1985....28P} van Paradijs, J. \ 1985, \aap, 144, 199
\bibitem[Warner (1995)]{1995....28P} Warner, B.\ 1995, Cataclysmic Variable Stars Cambridge Astrophysics Series (Cambridge: Cambridge Univ. Press), 28

\end{thebibliography}
\end{document}